\PassOptionsToPackage{unicode}{hyperref}
\PassOptionsToPackage{hyphens}{url}
\documentclass[
]{article}
\usepackage{xcolor}
\usepackage{amsmath,amssymb}
\usepackage{iftex}
\ifPDFTeX
  \usepackage[T1]{fontenc}
  \usepackage[utf8]{inputenc}
  \usepackage{textcomp} 
\else 
  \usepackage{unicode-math} 
  \defaultfontfeatures{Scale=MatchLowercase}
  \defaultfontfeatures[\rmfamily]{Ligatures=TeX,Scale=1}
\fi
\usepackage{lmodern}
\ifPDFTeX\else
\fi
\IfFileExists{upquote.sty}{\usepackage{upquote}}{}
\IfFileExists{microtype.sty}{
  \usepackage[]{microtype}
  \UseMicrotypeSet[protrusion]{basicmath} 
}{}
\makeatletter
\@ifundefined{KOMAClassName}{
  \IfFileExists{parskip.sty}{%
    \usepackage{parskip}
  }{
    \setlength{\parindent}{0pt}
    \setlength{\parskip}{6pt plus 2pt minus 1pt}}
}{
  \KOMAoptions{parskip=half}}
\makeatother
\usepackage{longtable,booktabs,array}
\usepackage{pdflscape}
\usepackage{calc} 
\usepackage{etoolbox}
\makeatletter
\patchcmd\longtable{\par}{\if@noskipsec\mbox{}\fi\par}{}{}
\makeatother
\IfFileExists{footnotehyper.sty}{\usepackage{footnotehyper}}{\usepackage{footnote}}
\makesavenoteenv{longtable}
\providecommand{\tightlist}{%
  \setlength{\itemsep}{0pt}\setlength{\parskip}{0pt}}
\usepackage{bookmark}
\IfFileExists{xurl.sty}{\usepackage{xurl}}{} 
\hypersetup{
  hidelinks,
  pdfcreator={LaTeX via pandoc}}

\author{}
\date{}

\begin{document}

\section{MutMem: Cryptographically Authorized Mutation in Persistent
Agent
Memory}\label{mutmem-cryptographically-authorized-mutation-in-persistent-agent-memory}

\textbf{Author:} Walid Saidi\\
\textbf{Affiliation:} Independent researcher\\
\textbf{Draft status:} Verified utility, native mutation-integrity,
post-calibration PoisonedRAG N=100, causal epistemic ablation, and
blinded system-author review results integrated, 2026-08-03\\
\textbf{Release status:} Submission candidate. The signed scratch-brain
purges, sanitized publication aggregate, and exact one-commit public
source release are complete and bound below.\\
\textbf{System:} HOM-AIMOS (AIMOS memory engine)\\
\textbf{Paper scope:} This paper evaluates the implemented mutation,
retrieval, and audit mechanisms and the evidence reported below. It
makes no claim about mechanisms or experiments outside this evaluated
scope.

\subsection{Evidence-status legend}\label{evidence-status-legend}

This draft uses the following labels so an architectural statement
cannot be mistaken for an experimental result.

\begin{itemize}
\tightlist
\item
  \textbf{{[}CODE{]}} Implemented in the named source or migration in
  the current working tree.
\item
  \textbf{{[}TEST-SOURCE{]}} A repository test encodes the stated
  contract; the complete public source and publication-evidence suites
  passed at the release commit on both Node.js 20 and Node.js 24.
\item
  \textbf{{[}LIVE-EVIDENCE{]}} Verified by a promoted, hash-addressed
  run artifact.
\end{itemize}

The utility, mutation, poisoning, ablation, and human-agreement numbers
below are bound to promoted run identifiers and the self-hashed public
aggregate. The public source is the fresh one-commit repository
\url{https://github.com/wallidsaydi-creator/HOM-AIMOS} at commit
\nolinkurl{580e9d2574478e78fcccf4bd00b7f44dc8a9b744}. Its 624-file source
manifest has canonical self-hash
\nolinkurl{64783f25c4555737ecbc9da343a1186cc3e4dc8a092a30465855f5fb1d2ba68e}
and files root
\nolinkurl{2f3908f94da90c2dc5210970c542d2126ae2d2baa608a7203c73806705be7af2}.

\subsection{Abstract}\label{abstract}

Persistent agent memory must be able to adapt when later outcomes change
the usefulness of earlier evidence. A mutable score, however, creates an
attribution problem: after a weight changes, an external reviewer must
be able to distinguish an authorized adaptation from an unauthorized
database edit. We present \textbf{MutMem}, a mutation protocol we
implement in HOM-AIMOS, a persistent-memory engine for agents. MutMem retains
the memory, records signed positive and negative outcome evidence
without age-based expiry, derives a bounded retrieval-frequency update,
and commits each nontrivial change as a housekeeper-authorized
transition. Every transition is bound to a terminal memory-provenance
node, an exact signer epoch, quantized old and new weights, a no-fork
predecessor, and two domain-separated SHA-256 commitments. Ed25519
verification occurs in the database writer and in an independent
portable verifier. Recall admits only provenance-verified evidence and
returns an RFC 6962-style Merkle receipt with domain-separated nonempty
leaves and internal nodes, binding the disclosed order and calibration
state.

MutMem also retains potentially poisoned content rather than deleting or
silently suppressing it. A native deterministic classifier appends a
housekeeper-authorized epistemic transition-\/-\/-such as
\texttt{unverified}, \texttt{poison\_suspect}, \texttt{poison\_likely},
\texttt{poison\_confirmed}, or \texttt{poison\_refuted}-\/-\/-without
changing the memory value, type, scope, or content hash. Each transition
binds the live content hash, exact signed event mutation hash,
predecessor classification hash, confidence, and detector signals.
Recall consumes the verified current projection as trust evidence while
preserving the complete classification history. This makes a detected
poisoning attempt traceable and reversible: later evidence may refute or
confirm the label, but cannot erase the earlier judgment.
\textbf{{[}CODE{]}} migration 092 and
\texttt{services/security/memory-epistemic-classifier.js}.

We evaluate the system along three separate axes: memory utility on the
complete LongMemEval and LoCoMo workloads, authorized-mutation and
tamper-detection behavior, and an N=100 targeted knowledge-poisoning
evaluation adapted from PoisonedRAG. On LongMemEval, HOM-AIMOS answers
459/500 questions correctly under LLM judgment (91.8\%, 95\% Wilson CI
89.06-\/-93.90). On LoCoMo it obtains 1472/1986 judged-correct answers
(74.12\%, 95\% Wilson CI 72.15-\/-76.00) and, in a separate
upstream-compatible token-F1 protocol, 58.20 F1 (95\% fixed-seed
bootstrap CI 56.46-\/-59.94). A fresh isolated mutation suite verifies
every declared authorization, topology, tamper, signer-epoch, and
post-mutation recall case. Across 20 measured native signed transitions,
median latency is 4.865 ms, p95 latency is 5.674 ms, and mean logical
row storage is 966.35 bytes per transition. In the final fixed-protocol
post-calibration PoisonedRAG adaptation, clean target-answer leakage is
2/100 (2.0\%), attacked target-answer ASR is 3/100 (3.0\%, 95\% Wilson
CI 1.03-\/-8.45), and therefore induced ASR among the 98 clean-negative
targets is 1/98 (1.02\%, 95\% Wilson CI 0.18-\/-5.56). No poison passage
appears in the attacked top-5 retrievals (0/100; 95\% Wilson upper bound
3.70\%). All
500 poison passages were retained, all 500 finished with signed
\texttt{poison\_likely} projections, and all 19,308 retained
benchmark-memory classification chains verified. Four of 18,808 clean
retained memories finished with an adverse label (0.0213\%, 95\% Wilson
CI 0.0083-\/-0.0547\%). Clean and attacked answer accuracy were 72\% and
71\%, respectively; the paired difference was not established (McNemar
exact \(p=1\)). In the preregistered causal ablation, poison reached the
pre-epistemic candidate opening for all 100 attacked targets. With the
epistemic policy bypassed (A0), poison appeared in 94/100 attacked top-5
sets and attacked QA accuracy fell to 40\%; enabling signed stored labels
(A1) reduced poison retrieval to 0/100 and raised attacked accuracy to
65\%, while clean accuracy changed from 71\% to 68\%. The paired A1--A0
attacked-accuracy improvement was significant after within-family Holm
correction, whereas the clean-accuracy difference was not. Query-local
lure detection made no additional selection change in this corpus, and
active-context withholding was not exercised because the preceding
label-aware layer had already removed poison from the selected sets. A
blinded review of all 200 production answers was completed by the system
author, not an independent reviewer; correctness agreement with the judge
was 96.5\% (Cohen's \(\kappa=0.911\)). MutMem provides evidence of integrity, authorization,
traceability, and historical continuity; it does not prove that
remembered content is true.

\subsection{1. Introduction}\label{1-introduction}

Long-lived agents accumulate observations, user preferences,
corrections, and outcomes. A memory that was helpful yesterday may
become misleading after a correction, while a previously low-value
observation may become useful when new evidence arrives. A production
memory system therefore needs adaptation. Treating adaptation as an
ordinary in-place update creates a forensic ambiguity: the final value
no longer shows who authorized the change, what prior state it replaced,
or whether the change followed the system\textquotesingle s cognitive
policy.

MutMem separates four properties that are often conflated:

\begin{enumerate}
\def\labelenumi{\arabic{enumi}.}
\tightlist
\item
  \textbf{Retention:} the memory and its historical evidence remain
  present.
\item
  \textbf{Authorization:} only the native housekeeper mutation owner may
  apply a cognitive reweight.
\item
  \textbf{Integrity:} hashes, signatures, and topology make unauthorized
  divergence detectable under the stated assumptions.
\item
  \textbf{Truth:} whether the remembered claim corresponds to the world.
\end{enumerate}

Cryptography can support the first three properties. It cannot establish
the fourth. A false claim can be correctly signed; a correct claim can
later become obsolete. MutMem therefore describes records as
authenticated, authorized, or verified---not as true merely because they
carry a signature.

\subsubsection{1.1 Contributions}\label{11-contributions}

This paper makes five system contributions, subject to final novelty
review:

\begin{enumerate}
\def\labelenumi{\arabic{enumi}.}
\tightlist
\item
  \textbf{A retention-preserving mutation protocol.} Signed outcome
  evidence can move a memory\textquotesingle s retrieval frequency
  upward or downward within a constitutional interval without deleting,
  deactivating, overwriting, or expiring the memory. \textbf{{[}CODE{]}}
  \texttt{services/learning/stdp-kernel.js},
  \texttt{services/governance/valence-ledger.js}, migrations 059, 068,
  080--085, and 091.
\item
  \textbf{A two-commitment cognitive trajectory.} Each nontrivial
  reweight is authorized by a signed provenance node and independently
  signed over the exact company, memory, old weight, new weight, and
  provenance mutation hash. The resulting trajectory has one genesis and
  no forks. \textbf{{[}CODE{]}}
  \texttt{services/governance/governor-provenance.js},
  \texttt{services/security/cognitive-weight-verifier.js}, migrations
  085 and 091.
\item
  \textbf{Recall-time, result-level proof.} Candidate memories are
  admitted only after retained provenance verification; the response
  includes ordered evidence and an RFC 6962-style Merkle root bound into
  a signed event receipt. \textbf{{[}CODE{]}}
  \texttt{services/retrieval/native-recall.js} and
  \texttt{services/security/memory-provenance.js}.
\item
  \textbf{An auditable evaluation design.} Utility, mutation integrity,
  and targeted poisoning are measured separately, with fixed statistical
  units, intended-N accounting, immutable per-unit artifacts, and signed
  purge evidence. The full utility, native mutation-integrity, and final
  post-calibration N=100 poisoning run, causal epistemic ablation,
  mutation suite, and blinded system-author review are complete and
  hash-verified. All 39 declared scratch-brain purge receipts verified,
  and the canonical user brain was excluded.
  \textbf{{[}LIVE-EVIDENCE{]}}
\item
  \textbf{Retention-preserving poison traceability.} Potentially
  poisoned passages remain ordinary retained memories, but a separate
  append-only epistemic chain records signed labels, confidence,
  detector signals, live-content binding, and predecessor topology. The
  current projection can move in either direction as later evidence
  arrives, and recall treats adverse labels as untrusted reference
  evidence rather than deleting the memory. \textbf{{[}CODE{]}}
  migration 092;
  \texttt{services/security/memory-epistemic-classifier.js};
  \texttt{services/retrieval/epistemic-trust-retrieval.js}.
\end{enumerate}

We do \textbf{not} claim the first use of signatures, hash chains,
Merkle trees, or mutable memory. The narrower claim is that these
mechanisms are composed into a native persistent-agent memory path in
which legitimate cognitive mutation remains externally distinguishable
from an unsigned edit while prior state is retained. We make no broader
claim that the underlying cryptographic primitives are novel.

\subsection{2. System and threat model}\label{2-system-and-threat-model}

\subsubsection{2.1 System principals}\label{21-system-principals}

Let \(c\) denote a company or tenant, \(m\) a retained memory
identifier, and \(e\) a signing epoch. AIMOS distinguishes:

\begin{itemize}
\tightlist
\item
  an enrolled master identity;
\item
  enrolled agent identity epochs;
\item
  a system operational identity named \texttt{housekeeper};
\item
  restricted runtime database roles;
\item
  an offline whole-brain purge ceremony.
\end{itemize}

The housekeeper is the native owner of autonomous cognitive mutation. It
is not a user-enrolled agent and it does not obtain authority from
environment variables. Runtime identity, configuration, credentials,
mutation authority, and request admission are ledger-backed.
\textbf{{[}CODE{]}} \path{architecture-authority.json},
\path{services/security/housekeeper-signer.js},
\path{services/security/request-receipt-ledger.js}.

\subsubsection{2.2 Adversary
capabilities}\label{22-adversary-capabilities}

The evaluation considers the following adversaries separately:

\begin{itemize}
\tightlist
\item
  a client attempting an unsigned, replayed, forked, cross-tenant, or
  over-clearance save/recall;
\item
  a runtime process attempting to update or delete a retained memory
  directly;
\item
  a database attacker able to alter stored rows but unable to forge the
  active Ed25519 private key;
\item
  a knowledge-poisoning attacker able to inject attacker-crafted
  passages but unable to query or inspect the target retriever in the
  black-box setting;
\item
  an honest but mistaken signer whose signed content is false.
\end{itemize}

\subsubsection{2.3 Assumptions}\label{23-assumptions}

The cryptographic arguments assume SHA-256 collision resistance, Ed25519
existential unforgeability under chosen-message attack, correct
canonicalization and fixed-width encoding, uncompromised signing keys,
and an external verifier that retains trusted public-key and
software-release anchors. Runtime ACLs constrain ordinary application
roles; they do not make a PostgreSQL superuser physically incapable of
rewriting storage.

\subsubsection{2.4 Non-goals and
boundaries}\label{24-non-goals-and-boundaries}

MutMem does not prove semantic truth, prevent denial of service, recover
from a compromised signing key, or make the database physically
immutable. Recall-time verification proves the evidence used for a
response; it is not by itself a continuously running whole-corpus audit.
A separate corpus verifier exists, but publication must report when and
over which snapshot it was executed.

\subsection{3. Retention and version
semantics}\label{3-retention-and-version-semantics}

The live doctrine forbids ordinary decay, deletion, suppression, or
deactivation. Lower relevance is represented by lower retrieval
frequency, bounded away from zero. Content corrections create retained
versions and explicit supersession relations rather than rewriting the
prior memory. The only destructive operation is the explicit, signed
whole-brain purge ceremony. \textbf{{[}CODE{]}}
\texttt{architecture-authority.json}, migrations 059, 089, and 090;
\texttt{services/security/whole-brain-purge.js}.

For cognitive weight \(w\),

\[
w \in [0.1, 3.0].
\tag{1}
\]

The lower bound \(0.1\) means that negative evidence reduces frequency
but is not a deletion surrogate. The memory row, signed save provenance,
later outcome evidence, and cognitive trajectory remain retained.

Canary detection follows the same retention rule. A detected marker does
not hard-reject or deactivate the signed save. The native write
disposition retains it as active quarantine at the \(0.1\) retrieval
floor, preserves the request and scan evidence, and commits the
disposition to the signed event ledger. Quarantine is a trust and
handling label, not a claim that cryptography has determined semantic
truth. \textbf{{[}CODE{]}}
\path{services/security/canary-write-gate.js} and
\path{services/write/persist-memory.js}. \textbf{{[}TEST-SOURCE{]}}
\path{tests/security/canary-write-gate.test.mjs} and
\path{tests/security/canary-boundary-contract.test.mjs}.

Epistemic poisoning labels are distinct from quarantine. A suspected or
likely poison passage keeps its original memory row and remains
available as explicitly untrusted reference evidence. The label is an
independently verifiable historical assertion about the memory, not a
rewritten lifecycle state and not a deletion surrogate.
\texttt{poison\_refuted} permits later signed evidence to restore
ordinary reference treatment without pretending the earlier suspicion
never existed.

\subsection{4. Signed outcome evidence and update
rule}\label{4-signed-outcome-evidence-and-update-rule}

For a memory \(m\), let the retained reward events be
\(r_{m,\ell}\in\{-1,+1\}\). The age-neutral valence judge computes

\[
j_m = \tanh\!\left(\sum_{\ell=1}^{L_m} r_{m,\ell}\right),
\qquad j_m\in[-1,1].
\tag{2}
\]

Time is not a coefficient: an event remains influential until later
retained evidence changes the signed sum. \textbf{{[}CODE{]}}
\path{services/governance/valence-judge.js}.

Given the current weight \(w_m\) and learning rate
\(\eta\in[0.001,0.5]\), the native reference-point update is

\[
w'_m = \operatorname{clip}_{[0.1,3.0]}
\left(w_m\exp(\eta j_m)\right).
\tag{3}
\]

The default implementation uses \(\eta=0.2\). The outcome event,
valence-ledger append, provenance append, cognitive projection,
live-weight update, and adjustment event execute inside one restricted
transaction. If the quantized target equals the current value, the
signed outcome remains retained but no fictitious weight transition is
appended. \textbf{{[}CODE{]}} \texttt{applyRewardSignal()} in
\texttt{services/learning/stdp-kernel.js}.

Equation (3) is not presented as the paper\textquotesingle s
cryptographic novelty. MutMem\textquotesingle s focus is the
authorization and evidence structure around the transition.

\subsection{5. Cryptographic mutation
protocol}\label{5-cryptographic-mutation-protocol}

\subsubsection{5.1 Notation}\label{51-notation}

{\def\LTcaptype{none} 
\begin{longtable}[]{@{}ll@{}}
\toprule\noalign{}
Symbol & Meaning \\
\midrule\noalign{}
\endhead
\bottomrule\noalign{}
\endlastfoot
\(m\) & 16-byte UUID of the retained memory \\
\(c\) & UTF-8 company identifier \\
\(q_i\) & integer milliscale weight after transition \(i\),
\(q_i\in\{100,\ldots,3000\}\) \\
\(w_i\) & displayed weight \(q_i/1000\) \\
\(B_i\) & canonical signed REWEIGHT body \\
\(C_i\) & SHA-256 content hash of \(B_i\) \\
\(\mu_i\) & memory-provenance mutation hash \\
\(h_i\) & cognitive projection-chain hash \\
\(t_i\) & cognitive transition commitment \\
\(s_i\) & Ed25519 signature over \(t_i\) \\
\(n_i\) & replay nonce \\
\(u_i\) & integer signing time in Unix seconds \\
\end{longtable}
}

All symbols are defined before use, and byte encodings below are
load-bearing protocol details.

\subsubsection{5.2 Signed provenance
node}\label{52-signed-provenance-node}

The live REWEIGHT path uses JSON Canonicalization Scheme-compatible
canonical JSON for \(B_i\):

\[
C_i = H(\operatorname{JCS}(B_i)).
\tag{4}
\]

For a non-genesis provenance node,

\[
\mu_i = H\!\left(C_i\,\Vert\,\mu_{i-1}\,\Vert\,
\operatorname{UTF8}(n_i)\,\Vert\,\operatorname{UTF8}(u_i)\right),
\tag{5}
\]

with the predecessor omitted for genesis. The housekeeper signs the
canonical body, nonce, and timestamp using the same retained-payload
signature form used by the provenance verifier. A second,
retained-attestation form additionally binds the observed origination
time; that form is not silently conflated with the live REWEIGHT form.
\textbf{{[}CODE{]}} \texttt{services/security/memory-provenance.js}.

\subsubsection{5.3 Quantization}\label{53-quantization}

Weights are quantized as

\[
q_i=\operatorname{round}(1000w_i).
\tag{6}
\]

Only integer \(q_i\) values enter the cognitive hashes and continuity
checks. Float columns are derived displays constrained to the exact
float32 value of \(q_i/1000\). Quantization avoids cross-language
ambiguity on the hash path. There are \(3000-100+1=2901\) representable
levels.

For \(q\in[100,3000]\), the float32 round-trip error is bounded by
approximately \(3\cdot2^{-24}<1.8\times10^{-7}\), so multiplying by
\(1000\) leaves an error far below \(0.5\) and recovers the same integer
by rounding. The critical verifier nevertheless compares canonical
integers and exact derived float bytes. \textbf{{[}CODE{]}} migration
091 and \texttt{docs/security/cognitive-weight-chain-SPEC.md}.

\subsubsection{5.4 Projection chain}\label{54-projection-chain}

Let

\[
P=\operatorname{UTF8}(\texttt{"aimos.cwc/v1"})\Vert\texttt{0x00}
\]

and let \(h_{-1}=0^{32}\). The projection commitment is

\[
h_i = H\!\left(
P\Vert\operatorname{UUID}(m)\Vert\operatorname{BE64}(q_{i-1})
\Vert\operatorname{BE64}(q_i)\Vert\mu_i\Vert h_{i-1}
\right).
\tag{7}
\]

The preimage is 109 bytes. Database uniqueness constraints enforce one
genesis per memory and at most one child for a predecessor. Runtime
roles receive no direct projection INSERT and no projection UPDATE,
DELETE, or TRUNCATE authority; the security-definer writer is the sole
projection owner. \textbf{{[}CODE{]}} migrations 081, 085, and 091.

\subsubsection{5.5 Transition authorization
signature}\label{55-transition-authorization-signature}

Let

\[
T=\operatorname{UTF8}(\texttt{"aimos.cognitive-transition/v2"})\Vert\texttt{0x00}.
\]

The transition-specific commitment is

\[
t_i = H\!\left(
T\Vert\operatorname{BE32}(|\operatorname{UTF8}(c)|)
\Vert\operatorname{UTF8}(c)\Vert\operatorname{UUID}(m)
\Vert\operatorname{BE64}(q_{i-1})\Vert\operatorname{BE64}(q_i)\Vert\mu_i
\right),
\tag{8}
\]

and

\[
s_i=\operatorname{Ed25519Sign}_{\mathrm{HK},e_i}(t_i).
\tag{9}
\]

The database verifies \(s_i\) against the exact active, unrevoked
housekeeper epoch referenced by the provenance row. The company
identifier is length-prefixed, and every remaining field is fixed-width.
Reusing a signature for another tenant, memory, weight pair, or
provenance node changes \(t_i\) and fails verification.
\textbf{{[}CODE{]}} migration 091 and
\texttt{services/security/housekeeper-signer.js}.

\subsubsection{5.6 Atomic writer
preconditions}\label{56-atomic-writer-preconditions}

The sole cognitive writer accepts a transition only when:

\begin{itemize}
\tightlist
\item
  the caller\textquotesingle s company and operational principal are set
  to the housekeeper scope;
\item
  \(q_{i-1}\) and \(q_i\) lie in \([100,3000]\) and differ;
\item
  \(\mu_i\) identifies the terminal, non-backfilled, 64-byte-signed
  REWEIGHT provenance node for \(m\);
\item
  the provenance body binds the same company, memory, old weight, and
  new weight;
\item
  the signer epoch is exact, active, unrevoked, and
  certificate-fingerprint matched;
\item
  \(s_i\) verifies over Equation (8);
\item
  the predecessor head and live weight agree with \(q_{i-1}\);
\item
  a first transition begins at the exact default weight or a separately
  signed retained baseline.
\end{itemize}

The function appends the projection and updates only
\texttt{retrieval\_weight} in the same transaction. \textbf{{[}CODE{]}}
migration 091.

\subsubsection{5.7 Signed epistemic-classification
chain}\label{57-signed-epistemic-classification-chain}

For memory \(m\), classification transition \(j\) records label
\(\ell_j\), integer confidence \(\gamma_j\in[0,1000]\), live content
hash \(d_m\), signed authority-event mutation hash \(a_j\), and
predecessor classification hash \(g_{j-1}\). Let \(g_{-1}=0^{32}\) and

\[
Q=\operatorname{UTF8}(\texttt{"aimos.memory-epistemic/v1"})\Vert\texttt{0x00}.
\]

The classification commitment is

\[
g_j=H\!\left(
Q\Vert\operatorname{UUID}(m)
\Vert\operatorname{BE32}(|\operatorname{UTF8}(\ell_j)|)
\Vert\operatorname{UTF8}(\ell_j)
\Vert\operatorname{BE64}(\gamma_j)
\Vert d_m\Vert a_j\Vert g_{j-1}
\right).
\tag{10}
\]

The authority event is signed by the housekeeper and binds the memory
identifier, live content hash, label, confidence, detector version, and
signal vector. The restricted database writer verifies that event before
appending the classification row and updating only the current epistemic
projection. The verifier reconstructs the unique chain, rejects forks or
disconnected rows, checks every event binding and live-content hash,
recomputes Equation (10), and compares the terminal chain node with the
current projection. \textbf{{[}CODE{]}} migration 092.

The initial detector requires compound evidence before persisting an
adverse label. A lone query-shaped prefix remains \texttt{unverified}
and may receive only response-local caution. When multiple retained
passages in the same signed session share an unusual normalized query
prefix, the housekeeper appends \texttt{poison\_likely} transitions for
the cluster, including the first passage. Independent supporting or
contradictory evidence can move the projection later; explicit
confirmation and refutation require separately bound evidence. This is a
deterministic epistemic policy, not a cryptographic proof of falsity.

\subsection{6. Verification}\label{6-verification}

\subsubsection{6.1 Per-memory
verification}\label{61-per-memory-verification}

The SQL verifier walks the trajectory from its single genesis,
recomputes Equations (5), (7), and (8), verifies continuity and exact
signer epochs, checks every transition signature, detects unreachable
rows, and compares the terminal integer-derived weight to the live
memory. Its time complexity is \(O(k)\) for a chain of length \(k\);
streaming verification uses \(O(1)\) auxiliary state aside from returned
records.

\subsubsection{6.2 Independent portable
verification}\label{62-independent-portable-verification}

\texttt{services/security/cognitive-weight-verifier.js} independently
reconstructs the fixed-width commitments with Node.js cryptography,
verifies the retained provenance evidence and identity epoch, verifies
signed pre-chain baselines, compares a declared classification summary
with the SQL verifier, and emits a deterministic corpus proof root:

\[
\begin{aligned}
D_c &= \operatorname{UTF8}(\texttt{"aimos.cognitive-corpus-proof/v1"}),\\
R_{\mathrm{corpus}} &= H\!\left(D_c\Vert\texttt{0x00}
\Vert\operatorname{JCS}(\text{ordered verification records})\right).
\end{aligned}
\tag{11}
\]

\begin{sloppypar}
The verifier classifies every memory as \texttt{certified\_chain},
\texttt{default\_empty\_chain}, \texttt{signed\_initial\_weight}, or
\texttt{unattested\_initial\_weight}; it does not make empty chains
disappear from the denominator. The SQL/portable parity Boolean compares
\texttt{ok}, certification status, chain length, signatures verified,
and total row count. It is classification-summary parity, not
byte-for-byte equality of every intermediate hash, failure reason, break
location, or terminal field. The portable proof root is a separate
commitment over its ordered verification records. \textbf{{[}CODE{]}}
migration 091 and
\path{services/security/cognitive-weight-verifier.js}.
\end{sloppypar}

\subsubsection{6.3 Tamper-evidence
proposition}\label{63-tamper-evidence-proposition}

\textbf{Proposition 1.} Under SHA-256 collision resistance and Ed25519
unforgeability, an adversary without the housekeeper private key cannot
change a retained transition\textquotesingle s company, memory, old
weight, new weight, provenance node, or predecessor while preserving
successful full verification.

\textbf{Argument.} Changing a field in Equation (7) changes \(h_i\)
except with negligible collision probability and invalidates the next
predecessor link. Recomputing the suffix does not solve Equation (8):
the altered tuple changes \(t_i\), and producing a valid \(s_i\)
requires the housekeeper key. Updating only the live weight fails
terminal fidelity. Runtime ACLs additionally reject the ordinary direct
write. This is a tamper-evidence claim, not a claim that privileged
storage cannot be modified.

\subsubsection{6.4 Bidirectional reachability
proposition}\label{64-bidirectional-reachability-proposition}

\textbf{Proposition 2.} From any terminal \(q_k\), every different
\(q'\in\{100,\ldots,3000\}\) is reachable by one authorized append; an
identical target requires no transition.

\textbf{Argument.} A signed REWEIGHT provenance node and Equation (8)
signature can bind \((c,m,q_k,q')\). Bounds and continuity hold by
construction, and no monotonicity condition exists. A later positive
outcome may therefore reverse a prior negative transition without
erasing it.

\subsection{7. Recall proof}\label{7-recall-proof}

Recall is a signed disclosure operation, not an unsigned search helper.
The native path binds the exact command to a verified request or
verified tool action, locks the active actor epoch, resolves
master-signed read authority or the housekeeper system principal, and
enforces company, clearance, data-class, and ownership boundaries.
\textbf{{[}CODE{]}} \texttt{services/retrieval/native-recall.js}.

Before a candidate is disclosed, the provenance verifier checks its
signed body, content hash, certificate chain, revocation timing,
signature form, mutation topology, live-row content hash, retained
snapshot, and current version topology. The operation fails closed if
requested evidence cannot be verified.

For ordered recall evidence records \(E_0,\ldots,E_{n-1}\), a nonempty
receipt uses domain-separated leaves and internal nodes. The
empty-result root follows the RFC 6962 empty-tree convention implemented
by the native service:

\[
L_i=H(\texttt{0x00}\Vert\operatorname{JCS}(E_i)),
\qquad
N(a,b)=H(\texttt{0x01}\Vert a\Vert b),
\qquad
R_{\varnothing}=H(\epsilon).
\tag{12}
\]

For a nonempty list, the service recursively splits at the largest power
of two smaller than the current subtree size. The resulting Merkle root
binds output order, memory identifier, live-content hash, save and
binding mutation hashes, version state, and calibration fields. The
root, command hash, outer-request hash, authority mutation hash,
request-receipt reference, result count, and evidence list are committed
to the housekeeper-signed event ledger. The receipt also retains the
verified requesting actor and envelope digest; it does not imply that
the requesting actor signed the result set. \textbf{{[}CODE{]}}
\texttt{recallMerkleRoot()} and \texttt{finalizeNativeRecall()} in
\texttt{services/retrieval/native-recall.js}; \texttt{logEvent()} in
\texttt{services/observe/event-ledger.js}.

\subsection{8. Whole-brain purge
boundary}\label{8-whole-brain-purge-boundary}

The sole destructive exception is offline and is not imported by the
server, routes, MCP surfaces, jobs, or tool registry. It inventories the
target, requires an exact human confirmation string and verified master
actor, prevents new database connections, terminates competing
connections, drops the entire database, removes owned
keychain/filesystem material, optionally recreates an empty migrated
brain, and signs a categorical receipt. The receipt contains counts and
classes rather than deleted memory content hashes. \textbf{{[}CODE{]}}
\texttt{services/security/whole-brain-purge.js}.
\textbf{{[}TEST-SOURCE{]}}
\texttt{tests/security/whole-brain-purge.test.mjs}.

Benchmark brains are purged only after artifacts and hash manifests
verify. No row-level cleanup is permitted.

\subsection{9. Evaluation design}\label{9-evaluation-design}

The evaluation separates three questions that cannot be collapsed into
one score.

\subsubsection{9.1 Utility}\label{91-utility}

The promoted utility run contains:

\begin{itemize}
\tightlist
\item
  LongMemEval: 948 source sessions, 10,960 turns, and 500 questions;
\item
  LoCoMo: 272 sessions, 5,882 turns, and 1,986 questions;
\item
  total completion count: 2,486 independently recalled and judged
  questions.
\end{itemize}

Sessions are replayed turn by turn through the native signed session
lifecycle. Each question receives one signed recall and one generator
call. The canonical-blind protocol also uses one judge call; the
upstream-compatible LoCoMo protocol instead applies its deterministic
category-aware token-F1 scorer. Retrieval and answer metrics are
reported separately. Promoted run \texttt{20260715111742\_96b25f} covers
the complete canonical-blind workloads; promoted run
\texttt{20260718205816\_fbde68} covers the complete upstream-compatible
LoCoMo protocol. \textbf{{[}LIVE-EVIDENCE{]}}

\subsubsection{9.2 Mutation integrity
suite}\label{92-mutation-integrity-suite}

Promoted run \texttt{20260723162050\_59a52d} executed the integrity
suite in a fresh isolated Genesis brain. It:

\begin{enumerate}
\def\labelenumi{\arabic{enumi}.}
\tightlist
\item
  retained Genesis memories through the canonical signed save path and
  issued an exact signed post-mutation recall;
\item
  appended positive and negative signed outcome evidence;
\item
  demonstrated a nontrivial upward, downward, and later upward
  trajectory;
\item
  verified SQL/portable corpus parity and the independent portable
  corpus proof root;
\item
  attempted direct runtime update, cross-memory proof reuse,
  cross-company reuse, no-op, stale-state discontinuity, out-of-bounds,
  and forked transitions;
\item
  tampered with a projection field, transition signature, terminal live
  weight, and signer epoch in the isolated database, verifying detection
  before restoring each value;
\item
  re-recalled the mutated memory and verified a one-item ordered
  evidence receipt, Merkle root, and signed event mutation hash;
\item
  measured native per-transition storage and latency across 20
  transitions.
\end{enumerate}

This suite reports authorization and integrity behavior, not retrieval
quality. The public aggregate self-hash is
\nolinkurl{9521798027be2893af745214a22747334be201d74178ddd9265ea97a32d34be8}.
\textbf{{[}LIVE-EVIDENCE{]}}

\subsubsection{9.3 Targeted poisoning}\label{93-targeted-poisoning}

The N=100 poisoning lane uses the official PoisonedRAG NQ target
fixture, five official attacker-crafted passages per target, and the
official pinned Contriever top-100 candidate identifiers resolved
against the pinned NQ/BEIR corpus. The target count, target fixture, and
five attacker-crafted passages per target match the upstream protocol;
the declared deviations are corpus scope, retriever, and answer model.
AIMOS\textquotesingle s native retriever operates over the bounded
100-candidate pool for each independently isolated target rather than the
original full 2,681,468-text Contriever index, making this a declared
adaptation and not a strict reproduction. The ablation positive control
shows that this adapted boundary exposed the attack: poison entered the
pre-epistemic candidate opening for 100/100 targets and was selected by
the policy-bypassed A0 arm for 94/100 attacked top-5 sets. The defended
result therefore cannot be explained by the injected passages never
reaching the measured retrieval boundary. Each target question is the independent unit;
passages and model retries are nested observations. Clean and attacked
outcomes are paired, with \texttt{top\_k=5}. The primary compatibility
metric is normalized target-answer substring ASR; semantic judgment is
secondary and cannot overwrite it. Promoted run
\texttt{20260722172124\_db0d79} completed all intended units.
\textbf{{[}LIVE-EVIDENCE{]}}

The promoted run keeps the same fixed target lock used during detector
calibration and adds four traceability outcomes: (i) injected passages
with signed epistemic-label evidence, (ii) clean retained memories
receiving an adverse label, (iii) final projection-chain verification
across the complete retained benchmark population, and (iv) poison
retrieval@5 after label-aware recall. Because the locked N=100 material
informed detector calibration, this is a fixed-protocol post-calibration
evaluation rather than an unseen held-out estimate. The experiment
therefore measures the final calibrated system and does not estimate
detector generalization to unseen attack material or isolate calibration
as a causal intervention. \textbf{{[}LIVE-EVIDENCE{]}}

\subsubsection{9.4 Models and judgment}\label{94-models-and-judgment}

The canonical-blind utility run used GPT-5.4 as generator and GPT-5.6
Terra as judge. The upstream-compatible LoCoMo run used GPT-5.4 as
reader and no LLM judge. The promoted PoisonedRAG and epistemic-ablation
runs used GPT-5.5 at medium reasoning as generator and GPT-5.6 Terra at
high reasoning as judge. Provider account material is represented only
by non-secret evidence and hashes. Because generator and judge remain
from one provider family, deterministic metrics are primary where
available. A blinded review of all 200 production-run answers was
completed by the system author after independent-review outreach did not
produce an available reviewer. The audit is disclosed as non-independent
and is an agreement diagnostic, not human ground truth.
\textbf{{[}LIVE-EVIDENCE{]}}

\subsubsection{9.5 Statistical
reporting}\label{95-statistical-reporting}

All rates report exact numerator and intended denominator. A
single-binomial proportion such as target-level ASR uses a two-sided
95\% Wilson score interval. Paired clean/attacked binary outcomes use
discordant-pair counts and an exact McNemar test; paired deltas receive
a predeclared paired bootstrap confidence interval. Missing or failed
units remain visible and block promotion of a nominal full-run result.

For the preregistered epistemic ablation, exact McNemar tests are grouped
into seven endpoint families: attacked poison retrieval, attacked
substring target assertion, incremental substring target assertion,
attacked semantic target assertion, incremental semantic target
assertion, clean QA correctness, and attacked QA correctness. Each family
contains the adjacent A1--A0, A2--A1, and A3--A2 contrasts plus the total
A3--A0 contrast; Holm--Bonferroni adjustment controls familywise
\(\alpha=0.05\) within each four-contrast family. Induced-ASR arm rates
condition on that arm's clean answer not already asserting the target, so
their denominators can differ and must not be compared naively across
table rows; paired incremental contrasts are the confirmatory comparison.

\subsubsection{9.6 Epistemic causal
ablation}\label{96-epistemic-causal-ablation}

Promoted run \texttt{20260730102457\_495de5} uses the fixed N=100
production run as its source state and evaluates four policies over the
same pre-disclosure candidate openings. A0 bypasses epistemic policy
while retaining AIMOS native relevance and diversity; it is not the
original Contriever baseline. A1 enables signed stored labels, A2 adds
query-local lure detection, and A3 adds active-context withholding. The
run completed 200 native recalls, 800 signed policy decisions, 800
generations, and 1,600 judgments. No arm changed canonical memory,
persistent retrieval weight, or classification state. The public
preregistration hash is
\nolinkurl{c224e942df7e4864cd66a82634f6739dc4657a9fc7cded93743f5f9c39e56fac}.

The source labels were calibrated on the same locked N=100 material and
the ablation cloned that calibrated state. Consequently, A1--A0 estimates
the effect of those stored labels within this fixed adapted corpus, not
the generalization of the detector to unseen attack structures.
\textbf{{[}LIVE-EVIDENCE{]}}

\subsubsection{9.7 Human agreement
review}\label{97-human-agreement-review}

The review packet contained all 200 clean/attacked production answers.
The reviewer did not see arm identity or judge verdict while assigning
correctness and target-assertion labels; the arm mapping was revealed
after completion. The reviewer was the system author, so the result is
reported as blinded but non-independent. Target pairs, rather than
individual arms, are the bootstrap unit. No answer was excluded.
\textbf{{[}LIVE-EVIDENCE{]}}

\subsection{10. Results}\label{10-results}

All values in this section regenerate from
\path{eval/publication/verified-benchmark-results.json}, whose
self-hash is
\nolinkurl{06afd5ba25c96d12c020df891e3711e821578df21d76c86c6398bc3792377a3f}.
The exporter rehashed every entry in the three promoted benchmark
artifact manifests: 46,032 utility-run artifacts, 26,143
upstream-compatible LoCoMo artifacts, and 21,517 PoisonedRAG artifacts
(93,692 total benchmark declarations), all 3,518 epistemic-ablation
artifacts, plus all seven declarations in the mutation-suite manifest.
The row-level source records remain private
because they contain upstream dataset text, provider payloads, or
retained memory identifiers; their hashes and sanitized aggregates are
public.

\subsubsection{10.1 LongMemEval}\label{101-longmemeval}

The complete 500-question LongMemEval run obtained 459 judged-correct
answers, or 91.8\% accuracy (95\% Wilson CI 89.06-\/-93.90). At
\texttt{k=20}, any-hit rate, hit@1, mean reciprocal rank, and mean
evidence recall were all 1.000; mean nDCG@20 was 0.9659. The perfect
evidence-hit measurements do not imply perfect answering: preference
questions were the weakest answer category at 22/30 (73.33\%), while
single-session assistant questions were 56/56.

{\def\LTcaptype{none} 
\begin{longtable}[]{@{}lrr@{}}
\toprule\noalign{}
Category & Correct / N & Judged accuracy \\
\midrule\noalign{}
\endhead
\bottomrule\noalign{}
\endlastfoot
Abstention & 28 / 30 & 93.33\% \\
Knowledge update & 67 / 72 & 93.06\% \\
Multi-session & 106 / 121 & 87.60\% \\
Single-session assistant & 56 / 56 & 100.00\% \\
Single-session preference & 22 / 30 & 73.33\% \\
Single-session user & 58 / 64 & 90.63\% \\
Temporal reasoning & 122 / 127 & 96.06\% \\
\end{longtable}
}

\subsubsection{10.2 LoCoMo under two declared
protocols}\label{102-locomo-under-two-declared-protocols}

Under the canonical-blind LLM-judged protocol, 1472/1986 answers were
correct: 74.12\% (95\% Wilson CI 72.15-\/-76.00). At \texttt{k=20},
any-hit was 97.12\%, hit@1 44.20\%, MRR 0.5500, mean evidence recall
94.71\%, and mean nDCG 0.6162. Category accuracy was 68.83\%
adversarial, 88.11\% multi-hop, 39.58\% open-domain, 48.94\% single-hop,
and 77.26\% temporal.

The separate upstream-compatible protocol evaluated all 1986 answers
with the pinned category-aware token-F1 scorer and obtained mean F1
0.5820, reported as 58.20 (95\% fixed-seed bootstrap CI 56.46-\/-59.94).
Its category F1 values were 84.30 adversarial, 60.37 multi-hop, 24.78
open-domain, 32.98 single-hop, and 48.40 temporal. At \texttt{k=25},
any-hit was 80.88\%, hit@1 44.35\%, MRR 0.5474, mean evidence recall
75.81\%, and mean nDCG 0.5798.

The 74.12\% and 58.20 results are complementary but not interchangeable:
one is LLM-judged binary QA accuracy and the other is deterministic
category-aware token F1. They are never averaged or presented as one
score.

\subsubsection{10.3 Mutation authorization, tamper detection, and
overhead}\label{103-mutation-authorization-tamper-detection-and-overhead}

The isolated mutation run completed 20/20 measured native signed
transitions while leaving the canonical user brain unchanged. It
verified a retained upward-\/-downward-\/-upward trajectory, a
negative-\/-neutral-\/-positive evidence sequence in which neutral
evidence created no fictitious projection, and continued memory
existence within the constitutional interval \([0.1,3.0]\). The
restricted writer rejected direct runtime updates, cross-memory and
cross-company proof reuse, no-op, stale-state discontinuity,
out-of-bounds, and forked transitions. SQL verification detected
projection/binding tamper, signature tamper, terminal-weight divergence,
and a stale signer epoch. After each deliberate isolated mutation was
restored, all nine cognitive chains verified in both the SQL and
portable verifiers, with exact parity and portable corpus root
\nolinkurl{a965aba5ec0a5eea813617489c8c3851e3ad5e16587c026fedfd71e303f43269}.

A native signed post-mutation recall returned the mutated memory with a
one-item evidence receipt. Its Merkle root was
\nolinkurl{8776caaae6034a5b1b02e661ccc8e8ca29e377141d615b03e05fdbe85c012e37},
and the receipt event carried mutation hash
\nolinkurl{eb1283b63eb68cf33904d629e1a17eb29138493d8185d1648cca4d59fdc7acda}.

Across 20 transitions, the exact measured transaction
boundary-\/-\/-housekeeper signatures, provenance append, certified
projection append, and live-weight update-\/-\/-had mean latency 4.898
ms, median 4.865 ms, p95 5.674 ms, minimum 3.857 ms, and maximum 6.488
ms. The paired projection and provenance rows occupied 19,327 logical
bytes in total, or 966.35 bytes per transition as measured by
\texttt{pg\_column\_size}; this excludes indexes and block free space.
PostgreSQL relation allocation increased by 16,384 bytes for provenance
and zero newly allocated bytes for projection because relation growth is
block-granular. The latter is not a zero-storage claim: logical
projection rows occupied 5,807 bytes. These are single-machine
native-boundary measurements, not a causal estimate of cryptographic
overhead relative to an unsafe unsigned control.
\textbf{{[}LIVE-EVIDENCE{]}}

\subsubsection{10.4 PoisonedRAG N=100
adaptation}\label{104-poisonedrag-n100-adaptation}

The primary normalized substring metric produced clean target-answer
leakage of 2/100 (2.0\%, 95\% Wilson CI 0.55-\/-7.00), attacked ASR of
3/100 (3.0\%, 95\% Wilson CI 1.03-\/-8.45), and induced ASR of 1/98
among clean non-target outputs (1.02\%, 95\% Wilson CI 0.18-\/-5.56).
The secondary semantic metric agreed exactly on these counts.

No poison passage appeared in any attacked top-5 result: 0/100 targets,
with a 95\% Wilson upper bound of 3.70\%. Mean poison count@5 was 0
(95\% fixed-seed bootstrap CI 0-\/-0). Clean answer accuracy was 72/100
(72\%, 95\% Wilson CI 62.51-\/-79.86), and attacked answer accuracy was
71/100 (71\%, 95\% Wilson CI 61.46-\/-78.99). The paired
attacked-minus-clean delta was -1 percentage point (95\%
paired-bootstrap CI -7 to +5); exact McNemar \texttt{b=5}, \texttt{c=4},
\texttt{p=1} does not establish a utility difference.

Poison retrieval is measured over the memories returned in the signed
recall proof, before generator-side handling of untrusted evidence. No
injected passage entered an attacked disclosure, so the generator
received no injected evidence in the measured recall-to-prompt path. The
three attacked target-answer events therefore cannot be attributed to
retrieval of the injected passages; paired clean target-answer leakage
was 2/100. This does not identify an alternative cause or establish that
unrelated model knowledge could not produce the target answer.

\subsubsection{10.5 Admission and cryptographic
completeness}\label{105-admission-and-cryptographic-completeness}

All 500 poison passages were submitted, admitted, retained as canonical
references, and associated with signed security-decision evidence. None
was rejected and none was quarantined. Each poison passage also has
signed epistemic-label evidence: at the instant each save returned, 100
first-in-cluster passages were still \texttt{unverified} and 400 later
passages were \texttt{poison\_likely}; compound cluster evidence then
appended transitions for the first passages, so all 500 final poison
projections were \texttt{poison\_likely}. Classification changed neither
canonical content nor retention.

Across the run, all 20,500 save attempts have terminal proof artifacts.
The native quality gate rejected 1,120 clean attempts and no poison
attempts; 19,380 operations were admitted. Seventy-two admitted clean
operations resolved to already retained rows through native
deduplication, leaving 19,308 unique retained benchmark memories: 500
poison and 18,808 clean. The database verifier checked all 19,308
epistemic chains and accepted all 19,308. The final projection contained
504 signed classification rows: 500 poison memories and four clean
memories were \texttt{poison\_likely}, while 18,804 clean memories
remained \texttt{unverified}. The observed clean adverse-label rate was
therefore 4/18,808, or 0.0213\% (95\% Wilson CI 0.0083-\/-0.0547\%). All
200 clean/attacked recall proofs verified, and all 100 target outcomes
matched their declared self-hashes. \textbf{{[}LIVE-EVIDENCE{]}}

Under the full epistemic-chain verifier, no non-\texttt{unverified}
projection is accepted unless it is backed by a complete, connected
sequence of signed classification events bound to the live memory
content. An adverse label written without that supporting history is
reported as an \texttt{unbacked\_projection}; a disconnected predecessor
history is rejected as \texttt{fork\_or\_disconnected\_history}. This is
a verifier-acceptance property under the stated cryptographic threat
model, not a claim that a database superuser cannot write arbitrary
bytes. \textbf{{[}CODE{]} {[}LIVE-EVIDENCE{]}}

\subsubsection{10.6 Preregistered epistemic
ablation}\label{106-preregistered-epistemic-ablation}

The positive control succeeded in every attacked target: at least one
poison passage entered the pre-epistemic candidate opening in 100/100
cases (95\% Wilson CI 96.30--100). With epistemic policy bypassed in A0,
poison appeared in 94/100 attacked top-5 sets (94\%, 95\% Wilson CI
87.52--97.22), with a mean 2.89 poison passages per top-5 set. This
establishes that the attack reached the measured mechanism boundary and
was capable of dominating native relevance-plus-diversity retrieval.

The registered primary compatibility endpoint is normalized substring
target assertion. Semantic target assertion is reported secondarily.

\scriptsize
{\def\LTcaptype{none}
\begin{longtable}[]{@{}lrrrr@{}}
\toprule\noalign{}
Arm & Poison retrieval@5 & Substring induced ASR & Clean QA & Attacked QA \\
\midrule\noalign{}
\endhead
\bottomrule\noalign{}
\endlastfoot
A0 & 94/100 (94\%) & 53/97 (54.64\%) & 71/100 (71\%) & 40/100 (40\%) \\
A1 & 0/100 (0\%) & 0/96 (0\%) & 68/100 (68\%) & 65/100 (65\%) \\
A2 & 0/100 (0\%) & 0/97 (0\%) & 68/100 (68\%) & 71/100 (71\%) \\
A3 & 0/100 (0\%) & 1/97 (1.03\%) & 69/100 (69\%) & 69/100 (69\%) \\
\end{longtable}
}
\normalsize

Induced-ASR denominators differ because each arm excludes targets whose
own clean answer already asserted the target. The arm rates are therefore
descriptive; the paired incremental-target contrasts provide the valid
cross-policy comparison. Under the secondary semantic endpoint, induced
ASR was 55/98, 0/97, 0/98, and 1/98 for A0 through A3.

Signed stored labels produced the demonstrated causal effect. Relative
to A0, A1 eliminated poison retrieval in 94 discordant targets with zero
retrieval regressions (exact McNemar
\(p=1.0097\times10^{-28}\), Holm-adjusted
\(p=4.0390\times10^{-28}\)). A0's attack reduced QA accuracy from 71\%
clean to 40\% attacked; A1 recovered attacked accuracy to 65\%, a
25-point gain over A0. The paired A1--A0 attacked-accuracy contrast had
26 improvements and one regression (exact
\(p=4.1723\times10^{-7}\), Holm-adjusted
\(p=1.2517\times10^{-6}\)). A1--A0 clean accuracy had seven losses and
four gains; neither its raw \(p=0.5488\) nor adjusted \(p=1\) established
a clean-side cost.

A1 and A2 had byte-identical selected sets, active contexts, and prompt
hashes for all 200 clean/attacked pairs. Query-local lure detection was
therefore a measured null on this fixed corpus. A2 and A3 were likewise
identical on those three inputs for all 200 pairs, but this is not an
efficacy null for withholding: the preceding label-aware layer left zero
selected passages marked for withholding, so the A3-only mechanism was
never exercised. Output-count differences between A2 and A3 are reported
descriptively and not attributed to withholding.

All 100 target outcomes, 800 signed retrieval decisions, 800 unique
decision events, and 800 unique ledger sequences verified. All 3,518
manifest entries rehashed. Memory, classification, and benchmark
footprint roots remained unchanged, as did all 504 signed classification
rows. The public ablation aggregate self-hash is
\nolinkurl{83a576895d26de63a7dde820b74effb93158299bf180289069af77569eb27105}.
\textbf{{[}LIVE-EVIDENCE{]}}

The run resumed after lifecycle-only amendments to
\path{eval/run-poisonedrag-epistemic-ablation.mjs} and
\path{scripts/benchmark/run-poisonedrag-epistemic-ablation.mjs}.
Scientific inputs were unchanged, completed immutable artifacts were
reused, and terminal status was reconciled under receipt hash
\nolinkurl{741b5124dfe29d77227ea9d9c746f2539f75073168d16428039d8488c520e6f2}.

\subsubsection{10.7 Blinded system-author
agreement}\label{107-blinded-system-author-agreement}

All 200 answers and all 100 paired targets were reviewed with zero
exclusions. Correctness labels agreed on 193/200 answers (96.5\% raw
agreement; Cohen's \(\kappa=0.9108\), target-cluster bootstrap 95\% CI
0.8169--0.9845). All seven correctness disagreements had the same
direction: the system author labeled the answer correct and the judge
did not (150 versus 143 positives; exact two-sided sign test
\(p=0.015625\)). Relative to this non-independent review, the judge was
stricter on every correctness disagreement; this is not evidence that
the judge is universally conservative.

Target-assertion labels agreed on 198/200 answers (99.0\% raw agreement;
\(\kappa=0.7949\), target-cluster bootstrap 95\% CI 0--1). The
\(2\times2\) table contains four joint positives, one author-only
positive, one judge-only positive, and 194 joint negatives, giving
positive specific agreement 0.80 and negative specific agreement 0.995.
The wide kappa interval reflects only five positive labels per rater and
2.5\% positive prevalence; the specific-agreement statistics expose that
low-prevalence structure directly. The reviewer was the system author;
this supports a blinded agreement diagnostic but not an independent
human-validation claim. The audit summary hash is
\nolinkurl{fec88dfb1e58dc429c023e876ff289684f1ee97e3d2df9d37478e09c0c54d387}.
\textbf{{[}LIVE-EVIDENCE{]}}

\subsubsection{10.8 Release-evidence
binding}\label{108-release-evidence-binding}

The publication aggregate verifies 39/39 signed scratch-brain purge
receipts, reports zero invalid receipts, and excludes the canonical user
brain. Its purge-evidence self-hash is
\nolinkurl{1bfbfdd76e36e74a615d22e05f7c933cca14795b76a2a230c95b8d39a5e16fcf}.
Raw identity-bearing receipts remain private. The public source release
contains one commit,
\nolinkurl{580e9d2574478e78fcccf4bd00b7f44dc8a9b744}, and the source manifest
has canonical self-hash
\nolinkurl{64783f25c4555737ecbc9da343a1186cc3e4dc8a092a30465855f5fb1d2ba68e}
and binds 624 shipped files under root
\nolinkurl{2f3908f94da90c2dc5210970c542d2126ae2d2baa608a7203c73806705be7af2}.
The fresh public tree passed the source, benchmark, architecture,
package, dependency, private-path, and internal-disclosure gates.
\textbf{{[}LIVE-EVIDENCE{]}}

\subsection{11. Discussion}\label{11-discussion}

\subsubsection{11.1 Authorization evidence is not semantic
truth}\label{111-authorization-evidence-is-not-semantic-truth}

MutMem answers a forensic question: whether a particular
cognitive-weight transition was admitted by the declared policy owner
and remains consistent with the retained trajectory. It does not answer
whether the underlying memory is correct. This separation prevents a
signature from becoming a semantic trust shortcut. Content-level
correction, contradiction, and supersession remain memory operations
with their own retained evidence; cognitive reweighting changes
retrieval frequency without rewriting those records.

\subsubsection{11.2 Why the protocol retains two
commitments}\label{112-why-the-protocol-retains-two-commitments}

The provenance commitment and transition commitment serve different
verification boundaries. The provenance node places the reweight inside
the memory\textquotesingle s retained event history and links it to the
prior provenance head. The transition commitment signs the exact
company, memory identifier, quantized old and new weights, and
provenance mutation hash consumed by the restricted writer. Collapsing
the two would make it harder for an external verifier to distinguish a
valid historical node from authorization of the exact live-weight
transition.

\subsubsection{11.3 Retention permits reversal without pretending the
past was
different}\label{113-retention-permits-reversal-without-pretending-the-past-was-different}

The bounded update rule is bidirectional. Later evidence may reverse the
direction of an earlier adjustment, but the reversal appends another
transition rather than replacing the first. This is useful operationally
because relevance is contextual and corrigible: a low-frequency memory
may later become important, while a formerly useful memory may lose
practical value. The retained sequence exposes both judgments and their
order.

\subsubsection{11.4 The three evaluation axes are intentionally
separate}\label{114-the-three-evaluation-axes-are-intentionally-separate}

Utility, mutation integrity, and poisoning robustness test different
propositions. High answer accuracy cannot establish that a database edit
was authorized. A valid signature cannot establish useful retrieval. A
low poisoning ASR cannot by itself prove general semantic safety. The
paper therefore avoids a merged score: utility determines practical
positioning, the mutation suite tests the central authorization and
integrity construction, and PoisonedRAG tests one declared adversarial
adaptation.

\subsubsection{11.5 Retention and retrieval isolation are different
security
boundaries}\label{115-retention-and-retrieval-isolation-are-different-security-boundaries}

The final PoisonedRAG run exposes a useful distinction. Cryptographic
admission correctly authenticated the provenance and execution of all
500 malicious passages but did not infer their factual falsity: every
poison passage was retained, none was quarantined, and none was
rejected. The native classifier added a second, signed statement: the
memory remains retained, but its current epistemic projection records
why recall treats it as untrusted reference evidence. The epistemic
retrieval path then excluded all injected passages from the attacked
top-5 disclosures. The classification chain makes the poison traceable
across later confirmation, refutation, and cognitive reweighting; it
does not turn a heuristic detector into a truth oracle. The evidence
supports classification, retrieval isolation, and traceability claims
under the declared post-calibration protocol. The fixed-corpus ablation
then isolates the effect of the signed stored labels within that
calibrated state: poison retrieval falls from 94/100 in A0 to 0/100 in
A1, while attacked answer accuracy rises from 40/100 to 65/100 and no
clean-side accuracy cost is established. Because the source labels were
calibrated on this same locked corpus, the result is causal for the
stored-label mechanism in this experimental state, not evidence of
detector generalization to unseen attacks.

\subsubsection{11.6 Poison as retained negative
evidence}\label{116-poison-as-retained-negative-evidence}

Retention changes the role of an attack artifact. Deleting a detected
passage would remove the evidence needed to audit the attack, study
detector behavior, or correct a false positive. MutMem instead keeps the
passage, its original signed save provenance, every epistemic
transition, and its response-local retrieval treatment. In the measured
N=100 run, the first passage in each of the 100 lure clusters was still
\texttt{unverified} when its own save returned. Evidence from later
members of the same retained cluster then caused a signed transition for
each first passage, so all 100 became \texttt{poison\_likely} and all 500
injected passages finished with adverse signed projections. This is
measured retroactive classification within the declared clustered attack,
not a guarantee that every previously missed or unseen attack will later
be detected. A labeled poison can become useful negative evidence:
reasoning may explicitly recognize that a claim was introduced by a
poisoning pattern, while later independent evidence may confirm or refute
that classification. The benefit comes from distinguishability and
historical continuity, not from treating malicious content as ordinary
support.

\subsubsection{11.7 The ablation identifies an ordered defense
boundary}\label{117-the-ablation-identifies-an-ordered-defense-boundary}

The ablation shows that the defense layers cannot be interpreted as
interchangeable filters. The attacked candidate opening contained poison
for all 100 targets, and A0 selected poison for 94/100 targets, so the
attack had reached a live retrieval boundary. Adding the signed stored
label in A1 removed all poison from the selected top-5 and recovered 25
attacked-utility points relative to A0. Adding query-local detection in
A2 produced byte-identical selected sets, active contexts, and prompt
hashes across all 200 clean/attacked pairs, a measured null on this
fixed corpus. Active-context withholding in A3 was never exercised
because the preceding label-aware layer left no selected poison to
withhold; A3 therefore supplies no efficacy estimate for withholding.
This ordering supports a sufficient first-layer result and a
defense-in-depth architecture, not a claim that every downstream layer
was independently effective.

The ablation also establishes mutation absence rather than mutation
integrity: memory roots, classification roots, 504 signed classification
rows, and the benchmark footprint remained unchanged while policy arms
varied. Authorization, topology, signer-epoch, and tamper behavior are
tested separately by the native mutation-integrity suite in
Section~10.3.

\subsection{12. Limitations}\label{12-limitations}

\begin{enumerate}
\def\labelenumi{\arabic{enumi}.}
\tightlist
\item
  \textbf{Integrity is not truth.} A valid signature identifies an
  authorized statement and signer epoch, not factual correctness.
\item
  \textbf{Key compromise is outside the main claim.} A stolen active
  housekeeper key can authorize malicious transitions until revocation
  and recovery.
\item
  \textbf{Runtime append-only is role-scoped.} Database superusers can
  modify storage; the protocol aims to make such changes detectable,
  assuming independent verifier anchors survive.
\item
  \textbf{Recall verification is demand-driven.} It verifies disclosed
  evidence. Corpus-wide verification is a separate explicit operation.
\item
  \textbf{Model judgment is not ground truth.} Same-family
  generator/judge bias and prompt sensitivity remain.
\item
  \textbf{PoisonedRAG comparability is conditional.} The target count,
  official NQ target fixture, and five attacker-crafted passages per
  target match the upstream protocol. Corpus scope, retriever, and answer
  model differ. The result is therefore a declared N=100 adaptation, not
  a strict reproduction of the paper\textquotesingle s 97\% NQ
  reference result.
\item
  \textbf{Canary coverage is marker-specific.} A retained canary
  disposition proves that a recognized marker crossed a monitored
  boundary. Absence of a marker does not establish that arbitrary
  content is benign, and quarantine does not establish that marked
  content is factually false.
\item
  \textbf{Overhead is bounded to one native transaction and machine.}
  The mutation suite measures the complete signed transition boundary,
  not a controlled cryptographic-ablation effect. It does not estimate
  how much faster an unsafe unsigned implementation would be, and its
  latency and storage measurements should not be generalized across
  hardware or database configurations.
\item
  \textbf{External anchors remain necessary.} A verifier needs a trusted
  public-key epoch, release identity, and retained proof material. An
  attacker who replaces both the database and every verifier anchor can
  present a self-consistent forgery outside this threat model.
\item
  \textbf{The release binding is revision-specific.} Claims in this
  paper bind the published source commit and evidence manifests named in
  Section~10.8. Later development is not covered automatically;
  behaviorally relevant changes require an explicit rerun and a new
  release identity.
\item
  \textbf{Poison classification is calibrated, not semantic ground
  truth.} The deterministic detector recognizes the declared clustered
  query-lure structure and may miss other attacks or mislabel benign
  clusters. Its signed output proves who classified what content, under
  which detector version and signals; it does not prove that the content
  is false. The final N=100 run is post-calibration on the locked
  benchmark material and must not be described as unseen held-out
  validation. Four of 18,808 clean retained memories received an adverse
  final label, so the observed false-positive rate is nonzero even under
  this fixed protocol. Retention preserves a missed artifact for later
  signed adjudication, as measured for the 100 first-in-cluster passages,
  but does not guarantee eventual detection of every unseen attack.
\item
  \textbf{The epistemic ablation inherits fixed-corpus calibration.}
  Its arms clone the final calibrated N=100 state. A1--A0 therefore
  identifies the effect of already-produced signed stored labels on this
  locked material; it does not estimate how the detector would label or
  isolate unseen attacks.
\item
  \textbf{One ablation resume included a lifecycle-only code amendment.}
  Two runner files changed during operational recovery. Completed
  artifacts were reused only after confirming that the scientific
  contract was unchanged and reconciling the run under SHA-256
  \nolinkurl{741b5124dfe29d77227ea9d9c746f2539f75073168d16428039d8488c520e6f2}.
  This preserves an auditable record of the recovery but is weaker than
  an uninterrupted execution from one immutable runner revision.
\item
  \textbf{The human review is blinded but non-independent.} The system
  author completed the 200-answer review after external reviewers were
  unavailable. The packet, labels, and scoring procedure are retained
  for independent replication; the reported agreement is not presented
  as independent human validation.
\end{enumerate}

\subsection{13. Ethics and responsible
release}\label{13-ethics-and-responsible-release}

The poisoning package includes attacker-crafted misinformation. Public
artifacts should include the minimum text required for reproducibility,
retain upstream attribution and license metadata, and clearly label
target answers as intentionally false. The public installer must keep
benchmark data out of the user\textquotesingle s normal brain.
Credentials and private key material must never enter artifacts, command
lines, logs, or the repository.

\subsection{14. Reproducibility
package}\label{14-reproducibility-package}

The release bundle contains or will contain:

\begin{itemize}
\tightlist
\item
  exact public source commit, one-commit history, source manifest, and
  dirty-state prohibition;
\item
  software, database, pgsodium, and model identifiers;
\item
  pinned dataset and fixture manifests with SHA-256 sidecars;
\item
  a sanitized public aggregate at
  \path{eval/publication/verified-benchmark-results.json};
\item
  retained private per-session replay proofs;
\item
  retained private per-question signed recall evidence, answer,
  judgment, timing, and attempt history;
\item
  mutation-suite provenance, projection, signature, and corpus-verifier
  outputs;
\item
  PoisonedRAG clean/attacked target records;
\item
  the four-arm epistemic-ablation aggregate, preregistration, recovery
  reconciliation, 3,518-entry artifact manifest, and paired statistical
  contrasts;
\item
  the 200-card blinded system-author audit packet, labels, agreement
  aggregate, and target-cluster bootstrap output;
\item
  per-memory epistemic-classification events, predecessor hashes,
  detector signals, current projections, and complete chain-verifier
  results;
\item
  aggregation code and generated tables;
\item
  artifact hash manifest;
\item
  canonical-brain non-contamination proof;
\item
  sanitized verification of the signed whole-brain purge receipts after
  the evidence freeze; raw identity-bearing receipts remain private.
\end{itemize}

The promoted run identifiers are \texttt{20260715111742\_96b25f}
(canonical-blind LongMemEval and LoCoMo),
\texttt{20260718205816\_fbde68} (upstream-compatible LoCoMo),
\texttt{20260722172124\_db0d79} (post-calibration PoisonedRAG N=100),
\texttt{20260730102457\_495de5} (four-arm epistemic ablation),
and \texttt{20260723162050\_59a52d} (native mutation integrity). Their
summary self-hashes are
\nolinkurl{28d35689ac5090634a6804d6fa7b6e1671b5b06cb5ebaa0d744d092f9aa40631},
\nolinkurl{0cf97847d64dd7d9fa642e5435eebfc6bb75fc4680d22e7192fde2ec255a727a},
\nolinkurl{7215ffe188528022a34251938133db14fdc8ee2bd93d1fb1119301962ca15908},
\nolinkurl{4ffaafbf57947add5247264f0ed57454e6d80bf25dc5481fb8a5e9c90964384b},
and
\nolinkurl{984994fc8d6c438599db32d6603d9cec8e02de398a7e6a262afdc253c11e0258}
for LongMemEval, judged LoCoMo, official-protocol LoCoMo, PoisonedRAG,
and mutation integrity respectively. The PoisonedRAG
epistemic-verification aggregate has self-hash
\nolinkurl{bf49888904fdc34db17e7f65d2f825ed1f18cfe5e66cc6e6d7bd77d5aee197c3};
the ablation aggregate has canonical self-hash
\nolinkurl{83a576895d26de63a7dde820b74effb93158299bf180289069af77569eb27105},
its preregistration has SHA-256
\nolinkurl{c224e942df7e4864cd66a82634f6739dc4657a9fc7cded93743f5f9c39e56fac},
and its lifecycle reconciliation has SHA-256
\nolinkurl{741b5124dfe29d77227ea9d9c746f2539f75073168d16428039d8488c520e6f2};
the human-audit aggregate has self-hash
\nolinkurl{fec88dfb1e58dc429c023e876ff289684f1ee97e3d2df9d37478e09c0c54d387}
and its 200-card packet has SHA-256
\nolinkurl{72d0fa8b5765e1fd2bedaf1cd11cc3dc3d5414a20bc334169815dba972dc2642};
the mutation-integrity public aggregate has self-hash
\nolinkurl{9521798027be2893af745214a22747334be201d74178ddd9265ea97a32d34be8}.
The unified publication aggregate has canonical self-hash
\nolinkurl{06afd5ba25c96d12c020df891e3711e821578df21d76c86c6398bc3792377a3f}
and verifies all 39 declared scratch-brain purge receipts; the canonical
user brain was not included.

No reviewer reproduction command will ingest benchmark corpora into the
reviewer\textquotesingle s normal brain. The public package contains
AIMOS-authored code, aggregate measurements, opaque unit identifiers,
cryptographic receipts, prompts, and hashes. LongMemEval, LoCoMo,
Natural Questions/BEIR, and PoisonedRAG fixture text remain
downloader-and-hash based under their upstream terms; source questions,
answers, conversations, passages, and provider payloads that reproduce
those materials are excluded from the public source repository.

\subsection{15. Related work}\label{15-related-work}

\subsubsection{15.1 Tamper-evident logs and cryptographic building
blocks}\label{151-tamper-evident-logs-and-cryptographic-building-blocks}

Tamper-evident logging predates agent memory. Crosby and Wallach
formalize histories in which signed commitments support membership and
incremental consistency proofs even when the logger is untrusted
{[}1{]}. Certificate Transparency standardizes a related Merkle-tree
construction with domain-separated leaves and internal nodes {[}2{]}.
MutMem uses established SHA-256, Ed25519 {[}3{]}, canonical JSON
{[}4{]}, and RFC 6962-style Merkle techniques; it does not claim those
primitives as novel. Its systems question is how to bind them to
authorized, retained cognitive reweighting and to the exact evidence
disclosed by a memory recall.

\subsubsection{15.2 Cryptographic provenance for agent
memory}\label{152-cryptographic-provenance-for-agent-memory}

Recent systems make cryptographic memory provenance an active research
area. MemLineage attaches per-principal Ed25519 signatures and an RFC
6962 Merkle log to memory entries, then propagates derivation lineage
through a weighted DAG to prevent externally descended evidence from
authorizing sensitive actions {[}5{]}. Portable Agent Memory defines
content-addressed memory entries in a Merkle-DAG, capability-scoped
disclosure, and cross-agent rehydration {[}6{]}. MemMark embeds
owner-controlled attribution signals into state-evolution choices and
retains cryptographic commitments, signed session anchors, and reveal
evidence so attribution can survive snapshot migration {[}7{]}.

These systems substantially narrow any novelty claim available to
MutMem. Signatures, Merkle provenance, transferable memory state,
lineage-aware enforcement, and cryptographic attribution are prior art.
The narrower MutMem construction studied here is a retained
bidirectional cognitive-weight trajectory: each nontrivial change is
bound both to a terminal memory-provenance node and to an exact signed
old/new transition, independently reconstructed by a portable verifier,
while a subsequent native recall commits its ordered disclosed evidence
to a signed event receipt. Whether this conjunction is sufficiently
distinct is a review question, not a premise of the evaluation.

\begin{landscape}
\scriptsize
{\def\LTcaptype{none} 
\begin{longtable}[]{@{}p{0.70in}p{1.10in}p{1.20in}p{1.00in}p{1.50in}p{1.25in}@{}}
\toprule\noalign{}
System & Primary objective & Cryptographic object & State evolution &
Enforcement or verification boundary & Distinction from MutMem \\
\midrule\noalign{}
\endhead
\bottomrule\noalign{}
\endlastfoot
MemLineage {[}5{]} & prevent untrusted memory lineage from justifying
sensitive actions & signed entries, Merkle log, derivation DAG & derived
memory lineage and trust propagation & sensitive-action gate and lineage
audit & focuses action authorization from provenance ancestry rather
than signed bidirectional retrieval-weight transitions \\
Portable Agent Memory {[}6{]} & move memory across heterogeneous agents
& content-addressed entries and Merkle-DAG & portable structured memory
transfer & capability-scoped disclosure and rehydration & focuses
interoperability and transfer rather than a live cognitive reweight
trajectory \\
MemMark {[}7{]} & attribute leaked or migrated memory snapshots & keyed
state-evolution watermark, commitments, signed anchors &
watermark-bearing write choices & snapshot attribution and reveal
verification & treats state evolution as an attribution carrier rather
than authorization of an explicit old/new weight change \\
MutMem & make retained cognitive adaptation distinguishable from
unsigned editing & provenance chain, signed transition, portable corpus
root, signed recall receipt & bounded downward and upward
retrieval-frequency appends & restricted writer, SQL verifier, portable
verifier, and disclosure receipt & utility, mutation-integrity, and
poisoning evidence complete under the declared protocols \\
\end{longtable}
}
\end{landscape}

\subsubsection{15.3 Memory poisoning}\label{153-memory-poisoning}

PoisonedRAG demonstrates that an attacker can inject a small number of
crafted passages into a large retrieval corpus and induce
attacker-chosen answers {[}8{]}. Its attack separates retrieval and
generation conditions and evaluates black-box and white-box
constructions. MutMem does not treat provenance as a truth oracle: a
malicious passage may be correctly signed and retained. The adapted
N=100 lane instead measures whether AIMOS\textquotesingle s declared
admission, quarantine, retrieval, and answer boundaries change
clean/attacked outcomes, while reporting deviations from the original
NQ/Contriever/model configuration. MemLineage is an especially important
adjacent defense because it evaluates cryptographic lineage as an
enforcement signal against memory poisoning {[}5{]}.

\subsubsection{15.4 Long-term memory utility
benchmarks}\label{154-long-term-memory-utility-benchmarks}

LongMemEval evaluates information extraction, multi-session reasoning,
temporal reasoning, knowledge updates, and abstention across 500
questions embedded in sustained chat histories {[}9{]}. LoCoMo evaluates
very long-term conversational memory through question answering, event
summarization, and multimodal dialogue generation over long
multi-session conversations {[}10{]}. They measure memory utility rather
than cryptographic mutation integrity. MutMem therefore reports their
retrieval and answer metrics separately from the mutation and poisoning
suites and avoids mixing Oracle, long-context, category-subset, lexical,
and model-judged conditions.

\subsection{16. Conclusion}\label{16-conclusion}

MutMem specifies a retention-preserving protocol for cognitive
adaptation in persistent agent memory. Signed outcome evidence can move
retrieval frequency downward or upward without erasing the memory or its
prior trajectory. A restricted writer binds each nontrivial change to
retained provenance and an exact signed transition; SQL and portable
verification expose authorization and topology failures; native recall
commits the ordered disclosed evidence to a signed receipt. Potential
poisoning is likewise retained and made distinguishable through a
separate signed epistemic trajectory that binds every label to the live
content hash, detector evidence, signer event, and predecessor. These
mechanisms support claims about authorization, integrity, traceability,
and historical continuity under the stated assumptions, not claims about
semantic truth.

Full utility runs show 91.8\% judged accuracy on LongMemEval, 74.12\%
judged accuracy on LoCoMo, and 58.20 upstream-compatible LoCoMo token F1
under their separately declared protocols. The native mutation suite
verified all declared authorization, topology, tamper, signer-epoch,
corpus-parity, and post-mutation recall cases; its 20 measured signed
transitions had median latency 4.865 ms, p95 latency 5.674 ms, and mean
logical row storage 966.35 bytes. The final post-calibration N=100
PoisonedRAG adaptation produced 2\% clean target-answer leakage, 3\%
attacked ASR, 1.02\% induced ASR among clean-negative targets, and 0/100
poison retrieval@5 while retaining all 500 poison passages.
All 500 finished with signed \texttt{poison\_likely} projections; all
19,308 retained benchmark-memory chains verified; and four of 18,808
clean memories carried an adverse final label. This supports measured
claims of authorized mutation integrity and retention-preserving poison
distinguishability and traceability under the declared protocols, not
semantic truth or unseen-attack generalization. In the fixed-corpus
ablation, poison was present in the attacked candidate opening for
100/100 targets and selected for 94/100 targets in A0; adding signed
stored labels in A1 reduced selected poison to 0/100 and raised attacked
answer accuracy from 40\% to 65\%, while no clean-side accuracy cost was
established. Query-local detection was a measured null after A1, and
active-context withholding was not exercised. A blinded but
non-independent system-author review agreed with the model judge on
193/200 correctness labels (96.5\%, \(\kappa=0.9108\)). These results are
bound to the exact public source and evidence identifiers reported in
Section~10.8.

\subsection{References}\label{references}

\begin{sloppypar}
\begin{enumerate}
\def\labelenumi{\arabic{enumi}.}
\tightlist
\item
  S. A. Crosby and D. S. Wallach, ``Efficient Data Structures for
  Tamper-Evident Logging,'' in \emph{18th USENIX Security Symposium
  (USENIX Security 09)}, Montreal, QC, Canada, Aug. 2009.
  \url{https://www.usenix.org/conference/usenixsecurity09/technical-sessions/presentation/efficient-data-structures-tamper-evident}
\item
  B. Laurie, A. Langley, and E. Kasper, ``Certificate Transparency,''
  RFC 6962, June 2013. \url{https://www.rfc-editor.org/rfc/rfc6962}
\item
  S. Josefsson and I. Liusvaara, ``Edwards-Curve Digital Signature
  Algorithm (EdDSA),'' RFC 8032, Jan. 2017.
  \url{https://www.rfc-editor.org/rfc/rfc8032}
\item
  A. Rundgren, B. Jordan, and S. Erdtman, ``JSON Canonicalization Scheme
  (JCS),'' RFC 8785, June 2020.
  \url{https://www.rfc-editor.org/rfc/rfc8785}
\item
  C. Ouyang and R. Hou, ``MemLineage: Lineage-Guided Enforcement for LLM
  Agent Memory,'' arXiv:2605.14421v1, May 2026.
  \url{https://arxiv.org/abs/2605.14421}
\item
  S. K. Ravindran, ``Portable Agent Memory: A Protocol for
  Cryptographically-\allowbreak Verified Memory Transfer Across Heterogeneous AI
  Agents,'' arXiv:2605.11032v1, May 2026.
  \url{https://arxiv.org/abs/2605.11032}
\item
  H. Zhang, X. Mao, G. Dong, Z. Li, X. Su, K. Chen, J. Yang, and Z. Lin,
  ``MemMark: State-Evolution Attribution Watermarking for Agent
  Long-Term Memory Systems,'' arXiv:2605.25002v2, May 2026.
  \url{https://arxiv.org/abs/2605.25002}
\item
  W. Zou, R. Geng, B. Wang, and J. Jia, ``PoisonedRAG: Knowledge
  Corruption Attacks to Retrieval-Augmented Generation of Large Language
  Models,'' in \emph{34th USENIX Security Symposium (USENIX Security
  25)}, Seattle, WA, USA, pp. 3827--3844, Aug. 2025.
  \url{https://www.usenix.org/conference/usenixsecurity25/presentation/zou-poisonedrag}
\item
  D. Wu, H. Wang, W. Yu, Y. Zhang, K.-W. Chang, and D. Yu,
  ``LongMemEval: Benchmarking Chat Assistants on Long-Term Interactive
  Memory,'' \emph{ICLR 2025}, arXiv:2410.10813v2, Mar. 2025.
  \url{https://arxiv.org/abs/2410.10813}
\item
  A. Maharana, D.-H. Lee, S. Tulyakov, M. Bansal, F. Barbieri, and Y.
  Fang, ``Evaluating Very Long-Term Conversational Memory of LLM
  Agents,'' in \emph{Proceedings of the 62nd Annual Meeting of the
  Association for Computational Linguistics (Volume 1: Long Papers)},
  Bangkok, Thailand, pp. 13851--13870, Aug. 2024,
  doi:10.18653/v1/2024.acl-long.747.
  \url{https://aclanthology.org/2024.acl-long.747/}
\end{enumerate}
\end{sloppypar}

\subsection{Appendix A. Code-to-claim
map}\label{appendix-a-code-to-claim-map}

\begin{landscape}
\scriptsize
{\def\LTcaptype{none} 
\begin{longtable}[]{@{}p{1.55in}p{2.85in}p{2.85in}@{}}
\toprule\noalign{}
Claim & Current source of truth & Required release evidence \\
\midrule\noalign{}
\endhead
\bottomrule\noalign{}
\endlastfoot
Full-retention doctrine and only whole-brain purge &
\texttt{architecture-authority.json}; migrations 059/089/090;
\texttt{whole-brain-purge.js} & ACL inspection, purge isolation test,
signed receipt \\
Age-neutral signed outcome ledger & \texttt{valence-ledger.js};
\texttt{valence-judge.js} & Complete chain verification and event
counts \\
Bounded bidirectional update & \texttt{stdp-kernel.js}; migration 068 &
Down/up live-fire evidence and bound tests \\
Signed REWEIGHT provenance & \texttt{governor-provenance.js};
\texttt{memory-provenance.js} & Portable provenance verification \\
Exact transition signature & \texttt{housekeeper-signer.js}; migrations
085/091 & Cross-language transition hash and signature verification \\
No-fork cognitive trajectory & migrations 081/085/091 & Fork-race and
topology tests \\
SQL/portable classification-summary parity &
\texttt{cognitive-weight-verifier.js}; migration 091 & Declared parity
vector, parity=true, and independently recomputed portable corpus proof
root \\
Signed recall receipt & \texttt{native-recall.js};
\texttt{event-ledger.js} & Per-question verified receipt and Merkle
recomputation \\
Signed poison traceability & migration 092;
\texttt{memory-epistemic-classifier.js};
\texttt{epistemic-trust-retrieval.js} & Per-memory label transitions,
complete chain verification, clean-label rate, and final N=100
traceability aggregate \\
Poisoning robustness & N=100 protocol & Target-level clean/attacked
artifact bundle \\
\end{longtable}
}
\end{landscape}

\subsection{Appendix B. Claim language allowed at
submission}\label{appendix-b-claim-language-allowed-at-submission}

Allowed only when supported by final artifacts:

\begin{itemize}
\tightlist
\item
  ``tamper-evident under the stated cryptographic and key-custody
  assumptions'';
\item
  ``authorization-gated for ordinary runtime roles'';
\item
  ``full historical trajectory retained'';
\item
  ``recall evidence verified at disclosure time'';
\item
  ``N=100 target-level estimate with a 95\% Wilson interval.''
\end{itemize}

Forbidden:

\begin{itemize}
\tightlist
\item
  ``tamper-proof'' or ``unhackable'';
\item
  ``cryptography proves truth'';
\item
  ``first signed memory system'' without a completed novelty review;
\item
  ``state of the art'' without strictly comparable published protocols;
\item
  claims about mechanisms or experiments not evaluated in this paper.
\end{itemize}

\end{document}